\documentstyle[12pt,epsfig]{article}
\begin{document}

\newcommand{\secm}{{\rm s}^{-1}}
\newcommand{\okgr}{{\Omega_{\rm K}}}
\newcommand{\eos}{equation of state~}
\newcommand{\eoss}{equations of state~}
\newcommand{\eossp}{equations of state}
\newcommand{\eosp}{equation of state}
\newcommand{\Eos}{Equation of state~}
\newcommand{\Eosp}{Equation of state}
\newcommand{\Eoss}{Equations of state~}
\newcommand{\Eossp}{Equations of state}
\newcommand{\mevt}{{\rm MeV/fm}^3}
\newcommand{\eps}{{\epsilon}}
\newcommand{\gsim}{{\stackrel{\textstyle >}{_\sim}}}
\newcommand{\lsim}{{\stackrel{\textstyle <}{_\sim}}}
\newcommand{\fm}{{\rm fm}}
\newcommand{\fmmo}{{\rm fm}^{-1}}
\newcommand{\fmmt}{{\rm fm}^{-3}}
\newcommand{\mev}{{\rm MeV}}
\newcommand{\msun}{M_{\odot}}
\newcommand{\ls}{{\stackrel{\textstyle <}{_\sim}}}
\newcommand{\gcmt}{{\rm g/cm}^3}
\newcommand{\el}{{\rm el}}
\newcommand{\pkgr}{{P_{\,\rm K}}}
\newcommand{\bag}{{B^{1/4}}}

\newcounter{sctn}
\newcounter{subsctn}[sctn]
\newcommand{\sctn}[1]{~\\ \refstepcounter{sctn} {\bf \thesctn~~ #1} \\ }
\newcommand{\subsctn}[1]
{~\\ \refstepcounter{subsctn} {\bf \thesctn.\thesubsctn~~ #1}\\}

\newcommand{\dateofdoc}{\today}

\newcommand{\tit}{\bf From Boson Condensation to Quark Deconfinement: \\
The Many Faces of Neutron Star Interiors}

\newcommand{\auth}{Fridolin Weber \\[6ex]}

\newcommand{\doe}{This work was supported by the Deutsche Forschungsgemeinschaft
  (DFG), by the Director, Office of Energy Research, Office of High Energy and
  Nuclear Physics, Division of Nuclear Physics, of the U.S. Department of
  Energy under Contract DE-AC03-76SF00098.}

\newcommand{\adr}{{Lawrence Berkeley National Laboratory \\ 
Nuclear Science Division, MS 70A-3307 \\
Berkeley, California 94720, USA \\ 
http://nta0.lbl.gov/$\sim$fweber} \\[4ex]}

\begin{titlepage}
\renewcommand{\thefootnote}{\fnsymbol{footnote}}
\setcounter{footnote}{0}
\begin{center}
\begin{Large}
\tit \\[7ex]
\end{Large}
\renewcommand{\thefootnote}{\fnsymbol{footnote}}
\begin{large}
\auth
\end{large}
\adr
\dateofdoc \\[30ex]
\end{center}

\begin{quote}
\begin{center} 
{Presented  at the \\
XXIII School of Theoretical Physics \\
Ustron, Poland  \\
15--22 September 1999 \\
To be published in the Acta Physica Polonica \\}
\end{center}
\end{quote}
\end{titlepage}

\begin{titlepage}
\tableofcontents
\end{titlepage}
\renewcommand{\thefootnote}{\arabic{footnote}}
\setcounter{footnote}{0}
\newpage

\renewcommand{\thefootnote}{\arabic{footnote}}
\setcounter{footnote}{0}
\begin{center}
\begin{Large}
\tit \\[4ex]
\end{Large}
\begin{large}
\auth
\end{large}
\adr
\end{center}
\vskip 1.0truecm

\begin{abstract}
  Gravity compresses the matter in the cores of neutron stars to densities
  which are significantly higher than the density of ordinary atomic nuclei,
  thus providing a high-pressure environment in which numerous particle
  processes -- from the generation of new baryonic particles to quark
  deconfinement to the formation of Boson condensates and H-matter -- may
  compete with each other. There are theoretical suggestions of even more
  `exotic' processes inside pulsars, such as the formation of absolutely stable
  strange quark matter, a configuration of matter even more stable than the
  most stable atomic nucleus, iron. In the latter event, neutron stars would be
  largely composed of pure quark matter, eventually enveloped in nuclear crust
  matter.  No matter which physical processes are actually realized inside
  neutron stars, each one leads to fingerprints, some more pronounced than
  others though, in the observable stellar quantities.  This feature combined
  with the tremendous recent progress in observational radio and X-ray
  astronomy, renders neutron stars to nearly ideal probes for a wide range of
  dense matter studies, complementing the quest of the behavior of superdense
  matter in terrestrial collider experiments.
\end{abstract}

\section{Introduction}\label{sec:intro}

Neutron stars are spotted as pulsars by radio telescopes and X-ray satellites.
They are more massive (i.e. $\sim 1.5\, \msun$) than our sun but are typically
only about $\sim 10$ kilometers across so that the matter in their centers is
compressed to densities that are up to an order of magnitude higher than the
density of atomic nuclei. A neutron star, therefore, provides a high-pressure
environment in which numerous subatomic particle processes are expected to
compete with each other and novel phases of matter -- like the quark-gluon
plasma being sought at the most powerful terrestrial particle colliders --
could exist. In my lecture, I will give an overview of the present status of
the research on the many phases of superdense matter in neutron stars, which
naturally is to be performed at the interface between nuclear physics, particle
physics and Einstein's theory of relativity.  Of particular interest will be
the existence of quark matter inside neutron stars and the fingerprints by
means of which this novel phase of matter could register itself in the observed
neutron star data. The detection of such matter in neutron stars would help to
clarify how quark matter behaves, and give a boost to theories about the early
Universe as well as laboratory searches for the production of quark matter in
heavy-ion colliders. Complementary talks on the physics of neutron stars will
be given by J.\ Lattimer and M.\ Prakash who will be discussing the structure
and evolution of neutron stars, and the neutrino interactions in dense matter,
respectively.

\section{The many faces of neutron stars}\label{sec:faces}

\begin{figure}[tb] 
\begin{center}
\leavevmode
\epsfig{figure=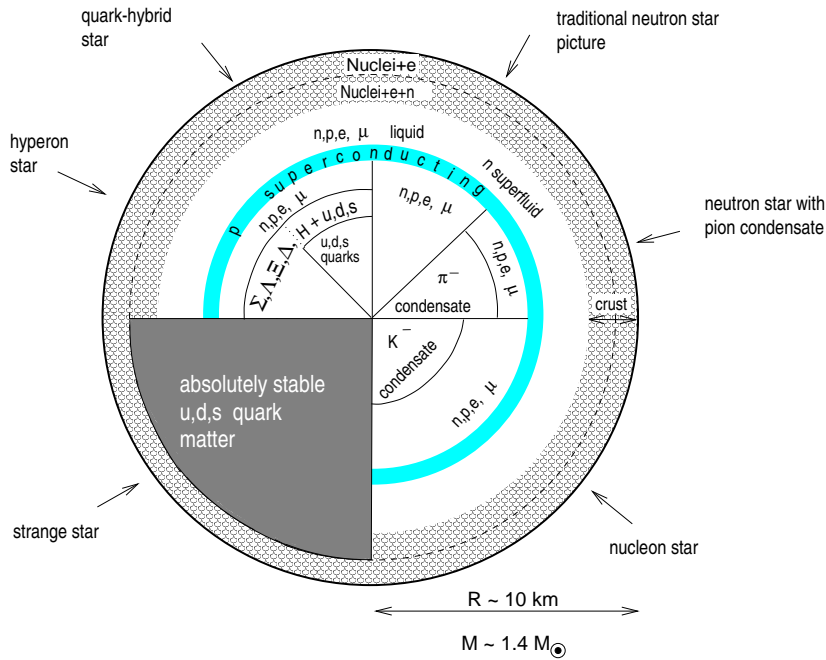,width=8.7cm,angle=-90}
\caption[]{Competing structures and novel phases of subatomic matter
  predicted by theory to make their appearance in the cores ($R\ls 8$~km) of
  neutron stars~\protect{\cite{weber99:book}}.}
\label{fig:cross}
\end{center}
\end{figure}

\begin{figure}[tb]
\begin{center}
\leavevmode 
\epsfig{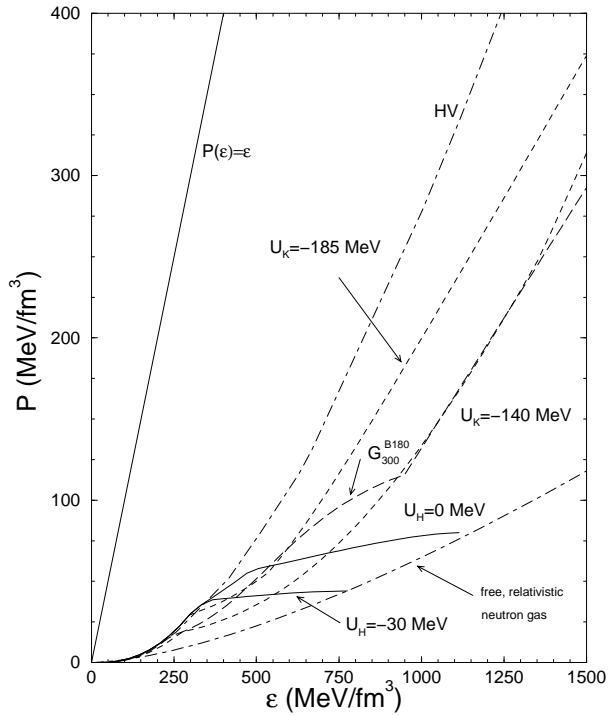} 
\caption[Graphical illustration of equations of state]{Models for the equation 
  of state (EoS) of `neutron' star matter
\protect{\cite{weber99:book,glen99:a,glen98:a}}.}
\label{fig:eos}
\end{center}
\end{figure}

From model calculations, it is known that neutron stars are far from being
composed of only neutrons but instead may possess rather complex interior
structures, as established in model calculations performed over the years.
Figure~\ref{fig:cross} gives an overview of those structures that are currently
most vividly discussed in the literature (for an overview, see
\cite{weber99:book}).
 No matter which physical structures are actually realized inside
neutron stars, each one leads to fingerprints, some more pronounced than others
though, in the \eos (i.e. pressure versus density) associated with these phases
of matter. This becomes very evident from Fig.~\ref{fig:eos},
which shows a collection of competing models for the \eos of neutron
star matter \cite{weber99:book}.  

\subsection{Hyperon star}

Only in the most primitive conception, a neutron star is constituted from
neutrons.  At a more accurate representation, a neutron star will contain
neutrons ($n$) and a small number of protons ($p$) whose charge is balanced by
leptons ($e^-$ and $\mu^-$). The interactions among the nucleons can be treated
in the framework of either Schroedinger-based theories
\cite{wiringa88:a,akmal98:a}, the semiclassical Thomas-Fermi method
\cite{strobel97:a}, or relativistic nuclear field theories solved at the
mean-field level \cite{weber99:book,glen97:book} or beyond
\cite{weber99:book,pal99:a,schertler99:a}. The coupling constants of the theory
must reproduce the bulk properties of nuclear matter at saturation density,
$\rho_0=0.16~\fmmt$ (energy density of $\eps_0 = 140~\mevt$).  These are the
binding energy $E/A$, effective nucleon mass $m^*_N/m_N$, incompressibility
$K$, and the symmetry energy $a_{\rm s}$ whose respective values are
\begin{equation}
  E/A=-16.0~\mev, ~ m^*_N/m_N=0.79, ~K\simeq 265~\mev, ~a_{\rm s}=32.5~\mev.
\label{eq:dcp.20}
\end{equation} Of the five, the value for the incompressibility of nuclear
matter carries some uncertainty.  Its value is currently believed to lie in the
range between about 180 and 300~MeV.  At the densities in the interior of
neutron stars, the neutron chemical potential, $\mu^n$, exceeds the mass
(modified by interactions) of various members of the baryon octet
\cite{glen85:b}. So in addition to nucleons and electrons, neutron stars may be
expected to have populations of hyperons, i.e. $\Sigma, \Lambda, \Xi$ and
eventually of $\Delta$'s.

\subsection{Nucleon stars}

Once the reaction 
\begin{equation}
  e^- \rightarrow K^- + \nu
\label{eq:kaon.1}
\end{equation} becomes possible in a neutron star, it becomes energetically
advantageous for the star to replace the fermionic electrons with the bosonic
$K^-$ mesons. Whether or not this actually happens depends on the mass of the
$K^-$ in dense matter. A handle on this is provided by the $K^-$ kinetic energy
spectra extracted from Ni+Ni collisions at SIS energies, measured by the KaoS
collaboration at GSI \cite{barth97:a}.
\begin{figure}[tb] 
\begin{center}
\leavevmode
\epsfig{figure=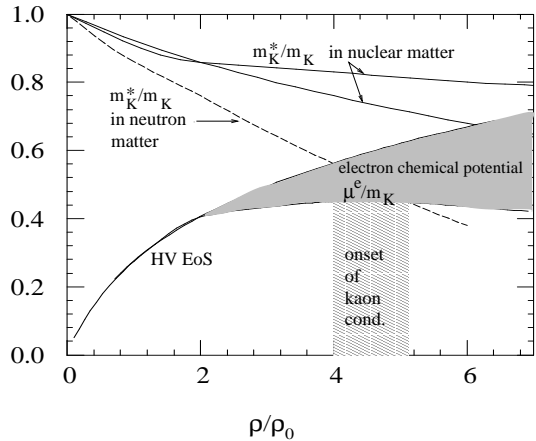,width=7.0cm,angle=0}
\caption[]{Effective kaon mass in nuclear \protect{\cite{mao99:a}} and neutron 
  star \protect{\cite{waas97:a}} matter.}
\label{fig:Kmass}
\end{center}
\end{figure} An analysis of the KaoS data shows that the attraction from
nuclear matter may bring the $K^-$ mass down to $m^*_{K^-}\simeq 200~\mev$ at
$\rho\sim 3\, \rho_0$. For neutron-rich matter, the relation
\begin{equation}
  m^*_{K^-}(\rho) \simeq m_{K^-}  \left( 1 - 0.2 \, {{\rho}\over{\rho_0}} \right)
\label{eq:meff02}
\end{equation} was established \cite{li97:a,li97:b,brown96:a,brown97:a}, with
$m_K = 495$~MeV the $K^-$ vacuum mass.  Values around $m^*_{K^-}\simeq
200~\mev$ lie in the vicinity of the electron chemical potential, $\mu^e$, in
neutron star matter \cite{weber99:book,glen85:b} so that the threshold
condition for the onset of $K^-$ condensation, $\mu^e = m^*_K$, which follows
from Eq.\ (\ref{eq:kaon.1}), could be fulfilled in the cores of neutron stars.
The situation is illustrated graphically in Fig.\ \ref{fig:Kmass}. Equation
(\ref{eq:kaon.1}) is followed by
\begin{equation}
  n + e^- \rightarrow p + K^- + \nu \, ,
\label{eq:npK.1}
\end{equation} with the neutrinos leaving the star. By this conversion the
nucleons in the cores of newly formed neutron stars can become half neutrons
and half protons, which lowers the energy per baryon of the matter
\cite{brown96:a}.  The relatively isospin symmetric composition achieved in
this way resembles the one of atomic nuclei, which are made up of roughly equal
numbers of neutrons and protons.  Neutron stars are therefore referred to, in
this picture, as nucleon stars.  The maximal possible mass of this type of star,
where Eq.\ (\ref{eq:npK.1}) has gone to completion, has been calculated to be
between about $1.5\, \msun$ \cite{thorsson94:a} and $1.8\, \msun$
\cite{glen99:a}.  Based on the former mass value, Brown et al.\ studied in a 
recent paper the formation and evolution of black holes in the Galaxy
\cite{brown99:a}.

Meson condensates can soften the \eos (EoS) considerably. As a consequence,
neutron stars with $K^-$ condensates can be rather dense and, therefore, have
radii smaller than neutron stars without condensates, i.e. $R \lsim 10$~km.  An
interesting candidate of such a small-radius object may be the nearby neutron
star RXJ~185~635--3754 \cite{walter96:a} whose radius could be as small as
$\sim 7$~km \cite{prakash97:bsky}.

\subsection{H-dibaryons}

A novel particle that could make its appearance in the center of a neutron star
is the so-called H-dibaryon, a doubly strange six-quark composite with spin and
isospin zero, and baryon number two \cite{jaffe77:a}. Since its first
prediction in 1977, the H-dibaryon has been the subject of many theoretical and
experimental studies as a possible candidate for a strongly bound exotic state.
In neutron stars, which may contain a significant fraction of $\Lambda$
hyperons, the $\Lambda$'s could combine to form H-dibaryons, which could give
way to the formation of H-matter at densities somewhere between $3\, \eps_0$
\cite{glen98:a} and $6\, \eps_0$ \cite{tamagaki91:a,sakai97:a}, depending on
the in-medium properties of the H-dibaryon. H-matter could thus exist in the
cores of moderately dense neutron stars.  In \cite{glen98:a} it was pointed out
that H-dibaryons with a vacuum mass of about 2.2~GeV and a moderately
attractive potential in the medium of about $- 30$~MeV could go into a Bose
condensate in the cores of neutron stars if the limiting star mass is about
that of the Hulse--Taylor pulsar PSR~1913+16, $M=1.444\, \msun$.  Conversely,
if the medium potential were moderately repulsive, around $+ 30$~MeV, the
formation of H-dibaryons may only take place in heavier neutron stars of mass
$M\gsim 1.6\, \msun$. If formed, however, H-matter may not remain dormant in
neutron stars but, because of its instability against compression could trigger
the conversion of neutron stars into hypothetical strange stars
\cite{sakai97:a,faessler97:a,faessler97:b}.

\subsection{Quark deconfinement in neutron stars}\label{ssec:deconf}

\begin{figure}[tb] 
\begin{center}
\leavevmode
\epsfig{figure=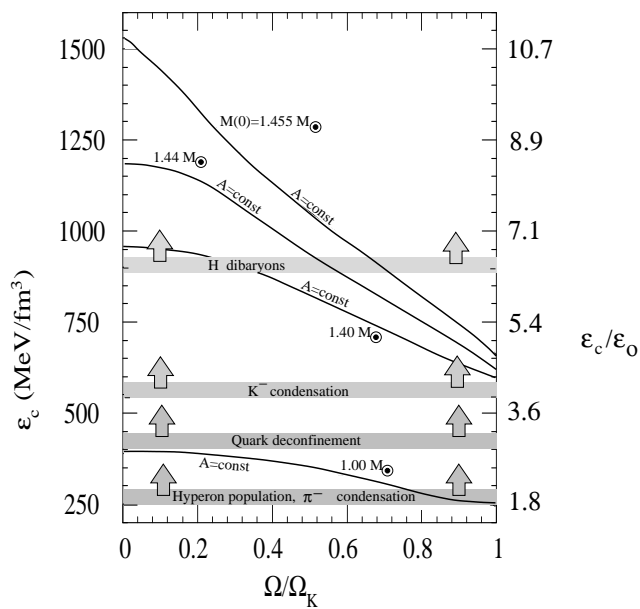,width=8.0cm,angle=-90}
\caption[]{Central density versus rotational frequency of
  sample neutron stars of constant baryon number, $A$.  Estimated threshold
  densities of various novel phases of superdense matter are indicated. $M(0)$
  is the non-rotating star mass,  $\okgr$ [cf.\ Eq.\ 
  (\protect{\ref{eq:okgr}})] stands for the Kepler frequency which 
  terminates stable rotation \protect{\cite{weber99:topr}}.}
\label{fig:ec1445fig}
\end{center}
\end{figure}

It has been suggested already back in the 1970's by a number of
researchers~\cite{fritzsch73:a,baym76:a,keister76:a,chap77:a,fech78:a,chap77:b}
that, because of the extreme densities reached in the cores of neutron
stars, neutrons protons plus the heavier constitutes ($\Sigma,
\Lambda, \Xi, \Delta$) can melt, creating the quark-gluon plasma state
being sought at the most powerful terrestrial heavy-ion colliders at
CERN [i.e. experiments NA35 NA44, NA45, CERES, and NA50; in a few years
at the LHC by the ALICE experiment] and RHIC. At present
one does not know from experiment at what density the expected phase
transition to quark matter occurs, and one has no conclusive guide yet
from lattice QCD simulations.  From simple geometrical considerations
it follows that nuclei begin to touch each other at densities of $\sim
(4\pi r^3_N/3)^{-1} \simeq 0.24~\fmmt$, which, for a characteristic
nucleon radius of $r_N\sim 1$~fm, is less than twice the baryon number
density $\rho_0$ of ordinary nuclear matter \cite{glen97:book}.  Above this
density, therefore, is appears plausible that the nuclear boundaries
of hadrons dissolve so that the formerly confined quarks now populate
free states outside of the hadrons.  Depending on rotational frequency
and stellar mass, densities as large as two to three times $\rho_0$
are easily surpassed in the cores of neutron stars, as can be seen
from Fig.\ \ref{fig:ec1445fig}, so that the neutrons and protons in
the centers of neutron stars may have been broken up into their
 constituent quarks by gravity \cite{weber99:topr}. More than that,
since the mass of the strange quark is only $m_s \sim 150$~MeV, high-energetic
up and down quarks are expected to readily transform to strange quarks at about
the same density at which up and down quark deconfinement sets in
\cite{glen91:pt,kettner94:b}.  Three flavor quark matter could thus exist as a
permanent component of matter in the centers of neutron stars
\cite{weber99:book,glen97:book,weber99:topr,glen97:a}. As we shall see in
Section~\ref{sec:signal}, radio astronomers may be able to spot evidence for
the existence of this novel phase of matter in the timing structure of pulsar
spin-down \cite{weber99:topr,glen97:a}.

\subsection{Diquark condensation and color superconductivity}

Very recently it was discovered that instantons may cause strong correlations
between up and down quarks, which could give way to the existence of colored
diquark pairs in superdense matter \cite{alford98:a,rapp98:a}.  These pairs
could form a Bose condensate in cold ($T < 50$~MeV) and dense ($\rho > 3\,
\rho_0$) quark matter.  Moreover, the condensate ought to exhibit color
superconductivity \cite{rapp98:a,rapp99:a}. Both the magnitude of the gap and
the critical temperature associated with the color superconductive phase were
estimated to be on the order of $\sim 100$~MeV \cite{rapp98:a,rapp99:a}!  The
implications of such tremendous gaps for the magnetic fields of pulsars and
their thermal evolution were explored in \cite{blaschke99:a,alford99:a} and
\cite{blaschke99:b}, respectively. Future theoretical studies of QCD at finite
baryon number density may reveal to which extent these newly established
features of quark matter will have their correspondence in a more complete
treatment of QCD at finite baryon number density.  They may also shed light on
whether or not a diquark condensate could possibly alter the \eos of neutron
star matter sufficiently strongly so that one may expect distinguishing
features in the global properties of neutron stars, too.

\subsection{Absolutely stable quark matter: the material of strange stars}
\label{ssec:ss}

So far we have assumed that quark matter forms a state of matter higher in
energy than atomic nuclei. This most `plausible' assumption, however, may be
quite deceiving \cite{bodmer71:a,witten84:a,terazawa89:a} because for a
collection of more than a few hundred $u,\, d,\, s$ quarks, the energy per
baryon ($E/A$) of quark matter can be just as well {\sl below} the energy of
the most stable atomic nucleus, $^{56}\rm{Fe}$, whose energy per baryon number
is $M(^{56}{\rm Fe})c^2/56=930.4$~MeV, with $M(^{56}{\rm Fe})$ the mass of the
$^{56}$Fe atom.  A simple estimate indicates that for strange quark matter $E/A
= 4 B \pi^2/ \mu^3$, so that bag constants of $B=57.5~\mevt$ (i.e.
$\bag=145$~MeV) and $B=85.3~\mevt$ ($\bag=160$~MeV) place the energy per baryon
of such matter at $E/A=829$~MeV and 915~MeV, respectively
\cite{weber99:book,madsen88:a,madsen93:a,madsen97:bsky}.  Obviously, these
values correspond to quark matter which is absolutely bound with respect to
$^{56}$Fe. In this event the ground state of the strong interaction would be
strange quark matter (strange matter), made up of $u,\, d,\, s$ quarks, instead
of nuclear matter.  This finding is one of the most startling predictions of
modern physics with far-reaching implications for neutron stars, for all
hadronic stellar configurations in Fig.\ \ref{fig:cross} would then be only
metastable with respect to a stars made up of absolutely stable 3-flavor
strange quark matter
\cite{weber99:book,witten84:a,alcock86:a,haensel86:a,alcock88:a}.  If this is
indeed the case, and if it is possible for neutron matter to tunnel to quark
matter in at least some neutron stars, then in appears likely that all neutron
stars would in fact be strange stars
\cite{madsen88:a,madsen93:a,madsen97:bsky,friedman90:a,caldwell91:a}.  In sharp
contrast to the other stars in Fig.\ \ref{fig:cross}, which are made up of
hadronic matter, possibly in phase equilibrium with quarks, strange stars
consist nearly entirely of pure 3-flavor quark matter, eventually
enveloped in a thin nuclear crust whose density is less than neutron drip
($4\times 10^{11}~\gcmt$) \cite{glen92:crust}.

The hypothetical, absolute stability of strange matter gives way to a variety
of novel stable strange matter objects which stretch from strangelets at the
small baryon number end, $A\sim 10^2$, to strange stars at the high end, $A\sim
10^{57}$, where strange matter becomes unstable against gravitational collapse
\cite{kettner94:b}.  The strange counterparts of ordinary atomic nuclei are the
strange nuggets, searched for in high-energy collisions at Brookhaven (e.g. 
E858, E864, E878, E882-B, E886, E896-A), CERN (Newmass experiment NA52),
balloon-borne experiments (CRASH), and terrestrial experiments (e.g. HADRON)
\cite{weber99:book}.  (For a recent review, see \cite{klingenberg99:topr}.)
Strange stars should possess properties that may allow one to distinguish them
from their `conventional'
counterparts \cite{weber99:book,glen97:book,alcock86:a,weber99:a,%
xu99:a,cheng96:a,usov98:a}.

\section{Stellar structure equations}

Since neutron stars are objects of highly compressed matter, the
geometry of spacetime is changed considerably from flat space. 
Neutron star models are thus to be constructed from 
Einstein's field equations of general relativity ($\mu, \nu$=0,1,2,3),
\begin{eqnarray}
  G^{\mu\nu} \equiv R^{\mu\nu} - {{1}\over{2}} g^{\mu\nu} R = 8 \pi
  T^{\mu\nu}(\eps,P(\eps)) \, ,
\label{eq:intro.1}
\end{eqnarray} which couples Einstein's curvature tensor, $G^{\mu\nu}$, to the
energy--momentum density tensor, $T^{\mu\nu}$, of the stellar matter. The
quantities $g^{\mu\nu}$ and $R$ in (\ref{eq:intro.1}) denote the metric tensor
and the Ricci scalar \cite{weber99:book}. Theories of superdense matter enter
in Eq.\ (\ref{eq:intro.1}) via $T^{\mu\nu}$, which contains the equation of
state, $P(\eps)$, of the stellar matter. It is derivable from a given
stellar-matter Lagrangian ${\cal L}(\{\phi\})$ \cite{weber99:book}.  In
general, ${\cal L}$ is a complicated function of the numerous hadron and quark
fields, collectively written as $\{ \phi \}$, that acquire finite amplitudes up
to the highest densities reached in the cores of compact stars.  According to
what has been said in Section~\ref{sec:faces}, plausible candidates for $\phi$
are the charged states of the SU(3) baryon octet, $p, n, \Sigma, \Lambda, \Xi$
\cite{glen85:b}, the charged states of the $\Delta$ \cite{weber89:e,huber98:a},
$\pi^-$ \cite{umeda92:a} and $K^-$
\cite{glen99:a,li97:a,li97:b,brown96:a,brown97:a} mesons, as well as the $u, d,
s$ quark fields \cite{kettner94:b}. The conditions of chemical equilibrium and
electric charge neutrality of the stellar matter require the presence of
leptons too, in which case $\phi=e^-, \mu^-$.  Models for the \eos then follow
according to the scheme (for details, see \cite{weber99:book})
\begin{eqnarray}
  { {\partial {\cal L}(\{\phi\}) }\over{\partial \phi} } -
    \partial_\mu \; { {\partial {\cal L}(\{\phi\})}\over{\partial
    (\partial_\mu \phi)} } = 0 \quad \Rightarrow \quad P(\eps) \, .
\label{eq:intro.2}
\end{eqnarray} 
In general, Eqs.\ (\ref{eq:intro.1}) and (\ref{eq:intro.2}) were to be
solved simultaneously since the particles move in curved spacetime
whose geometry, determined by Einstein's field equations, is coupled
to the total mass energy $\eps$ of the matter.  In the case of
neutron stars, however, the long-range gravitational forces can be
cleanly separated from the short-range forces, so that Eqs.\
(\ref{eq:intro.1}) and (\ref{eq:intro.2}) constitute two decouple problems.

\subsection{Non-rotating stars}

The structure equation of spherical neutron stars has been derived
from Einstein's equation (\ref{eq:intro.1}) first by Tolman
\cite{tolman39:a}, and Oppenheimer and Volkoff \cite{oppenheimer39}.
It reads
\begin{eqnarray}
  {{dP}\over{dr}} = - \, \frac{\epsilon(r) m(r)} {r^2} \frac{ \left( 1
+ P(r)/\epsilon(r) \right) \left( 1 + 4 \pi r^3 P(r)/m(r) \right) } {1
- 2 m(r)/r} \; ,
\label{eq:f28}
\end{eqnarray} and is known in the literature as the Tolman-Oppenheimer-Volkoff
equation, applicable to stellar configurations in hydrostatic equilibrium. We
use units for which the gravitational constant and velocity of light are
$G=c=1$ so that the mass of the sun is $\msun = 1.47$ km. The mass $m(r)$
contained in a sphere of radius $r$ is given by $m(r) = 4 \pi \int^r_0 r^2
\epsilon(r) dr$. Hence the star's total mass follows as $M\equiv m(R)$.

\subsection{Rotating stars}

The stellar equations describing rotating compact stars are
considerably more complicated than those of non-rotating compact stars
\cite{weber99:book}.  These complications have their cause in the
deformation of rotating stars plus the general relativistic effect of
the dragging of local inertial frames. This reflects itself in a metric
of the form \cite{weber99:book,friedman86:a}
\begin{eqnarray}
  {\rm d} s^2 = - \, {\rm e}^{2\,\nu} \, {\rm d} t^2 + {\rm e}^{2\,\psi} \,
  \bigl( {\rm d} \phi - \omega \, {\rm d} t \bigr)^2 + {\rm e}^{2\,\mu} \,
  {\rm d}\theta^2 + {\rm e}^{2\,\lambda} \, {\rm d} r^2 \, ,
\label{eq:f220.exact} 
\end{eqnarray} where each metric function, i.e. $\nu$, $\psi$, $\mu$ and
$\lambda$, depends on the radial coordinate $r$, polar angle $\theta$, and
implicitly on the star's angular velocity $\Omega$.  The quantity $\omega$
denotes the angular velocity of the local inertial frames, which are dragged
along in the direction of the star's rotation. This frequency too depends on
$r$, $\theta$ and $\Omega$.  Of particular interest is the relative frame
dragging frequency $\bar\omega$ defined as $ \bar\omega(r,\theta,\Omega) \equiv
\Omega - \omega(r,\theta,\Omega)$, which typically increases from about 15\% at
the surface to about 60\% at the center of a neutron star that rotates at its
Kepler frequency \cite{weber99:book,weber99:topr}. The Kepler frequency,
$\okgr$, is the maximum frequency a star can have before mass loss (mass
shedding) at the equator sets in. It sets an absolute upper limit on stable
rapid rotation. In classical mechanics the expression for $\okgr$, determined
by the equality between centrifuge and gravity, is readily obtained as $\okgr =
\sqrt{M/R^3}$. Its general relativistic counterpart is given by
\cite{weber99:book,friedman86:a}
\begin{eqnarray}
  \okgr = \omega +\frac{\omega^\prime}{2\psi^\prime} +e^{\nu -\psi}
    \sqrt{ \frac{\nu^\prime}{\psi^\prime} +
    \Bigl(\frac{\omega^\prime}{2 \psi^\prime}e^{\psi-\nu}\Bigr)^2 } \,
    , \qquad \pkgr \equiv {{2 \pi} \over {\okgr}} \, .
\label{eq:okgr}  
\end{eqnarray} The primes denote derivatives with respect to the
Schwarzschild radial coordinate.  Equation (\ref{eq:okgr}) is to be
evaluated self-consistently together with Einstein's field equations
(\ref{eq:intro.1}) for a given model for the \eosp.

\begin{figure}[tb]
\begin{center}
\leavevmode 
\epsfig{figure=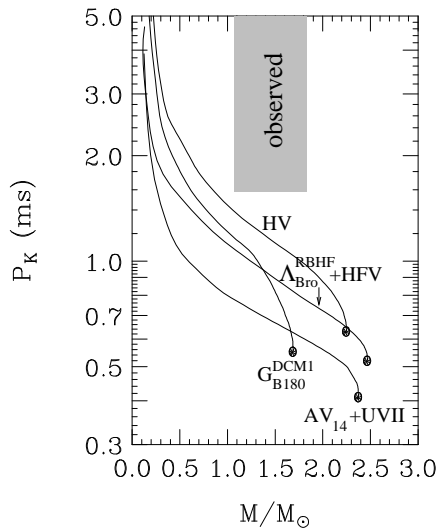,width=7.0cm,angle=90}
\caption[Kepler period versus rotational mass]{Onset of mass shedding
from rapidly spinning neutron stars, computed for a sample of \eoss
\protect{\cite{weber99:book}}. The Kepler period is defined in Eq.\
(\ref{eq:okgr}).}
\label{fig:1}
\end{center}
\end{figure}

Figure~\ref{fig:1} shows $\okgr$ as a function of rotating star mass. The
rectangle indicates both the approximate range of observed neutron star masses
as well as the observed rotational periods which, currently, are $P \geq 1.6$
ms.  One sees that
 all pulsars so far observed rotate below the
mass shedding frequency and so can be interpreted as rotating 
neutron stars. Half-millisecond periods or even smaller ones
are completely excluded for neutron stars of mass $1.4\, \msun$
\cite{friedman86:a,friedman89:a,sato89:a} for these \eoss
\cite{weber99:book}. The situation appears to be very different for
neutron stars made up of self-bound strange quark matter, the
so-called strange stars introduced in Section~\ref{ssec:ss}.  Such
stars appear to withstand stable rotation against mass shedding down
to rotational periods in the half-millisecond regime or even less
\cite{glen91:a}.  As a consequence, the possible future discovery of a
single sub-millisecond pulsar, say 0.5~ms, could give a strong hint
that, firstly, strange stars actually exist and, secondly, the
deconfined, self-bound phase of 3-flavor strange quark matter is the
true ground state of the strong interaction rather than nuclear
matter.  This conclusion is strengthened by the finding of
\cite{madsen98:a} that young strange stars appear not to be subject to
the recently discovered $r$-mode instability, which would slow down
hot neutron stars to periods of several milliseconds via the emission
of gravitational radiation within a year after birth.
 
\section{Mass constraints from QPOs in LMXBs}

\begin{figure}[tb]
\begin{center}
\leavevmode
\epsfig{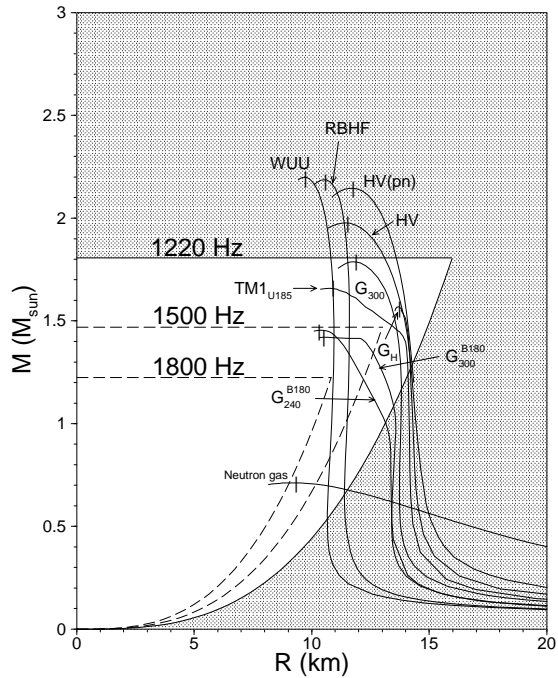}
\caption[Neutron star mass versus radius]{Neutron star mass versus
  radius for several \eoss shown in Figure \ref{fig:eos}.  Constraints derived
  from the observation of a QPO frequency of 1220~Hz in neutron star
  4U\,1636--536 are indicated by the shaded area \protect{\cite{miller98:a}}.}
\label{fig:mrad2}
\end{center}
\end{figure}

Figure \ref{fig:mrad2} exhibits the mass of non-rotating neutron stars
as a function of star radius.  Each stellar sequence is shown up to
the maximum-mass star, indicated by tick marks. Stars beyond the mass
peak are unstable against radial oscillations and would collapse to
black holes. Evidently, all \eoss can accommodate a neutron star as heavy
as the Hulse-Taylor pulsar, whose mass is very precisely known to be
$M({\rm PSR}~1913+16) = 1.444 \, \msun$.
 Rather heavy neutron stars, on the other had, of mass $M\sim
2\,M_\odot$ can only be obtained for \eoss that exhibit a rather stiff behavior
at supernuclear densities (cf.\ Fig.\ \ref{fig:eos}).  

Knowledge of the maximum-mass value is of great importance for two reasons.
Firstly, about 20 neutron stars masses are presently known \cite{thorsett99:a},
and the largest of these imposes a lower bound on the maximum mass of a
theoretical model. The current lower bound on the maximum-mass is about
$1.56\,\msun$ (i.e. neutron star 4U\,0900--40), which, if firmly established,
would indicate that the \eos of superdense matter will be rather soft at
supernuclear densities.  This would change if the maximum-mass should be closer
to the upper bound of $1.98\,\msun$. Indications for the possible existence of
such heavy neutron stars, with masses around $2\,\msun$, may come from the
observation of quasi-periodic oscillations (QPOs) in luminosity in low-mass
X-ray binaries (LMXBs) \cite{miller98:a,bildsten95:a,kaaret97:a,miller98:b}.
If confirmed, a significant fraction of \eoss presently discussed in the
literature \cite{weber99:book,schaab99:a} could be ruled out.  Because of the
significant stiffness of the EoS required by heavy neutron stars, all the phase
transitions reviewed in Section \ref{sec:faces}, which imply generally a
softening of the EoS rather than a stiffening, appear to be unfavored if not
completely ruled out.  In this event neutron star matter were likely to be made
up of chemically equilibrated nucleons only \cite{schaab99:a}.  Rapid neutron
star rotation increases the non-rotating maximum mass value by at most 25\%
\cite{weber99:a,friedman86:a,salgado94:a,cook94:a,cook94:b,eriguchi94:a}, so
that even extremely rapidly spinning neutron stars can not have masses
significantly above $2.5\, \msun$, as can be seen from Fig.\ \ref{fig:1}.  The
second reason is that the maximum mass can be useful in identifying black hole
candidates \cite{ruffini78:a,brown94:a,bethe95:a}.  For example, if the mass of
a compact companion of an optical star is determined to exceed the maximum mass
of a neutron star, it must be a black hole.  Because the
\begin{figure}[tb]
\begin{center}
\leavevmode
\epsfig{figure=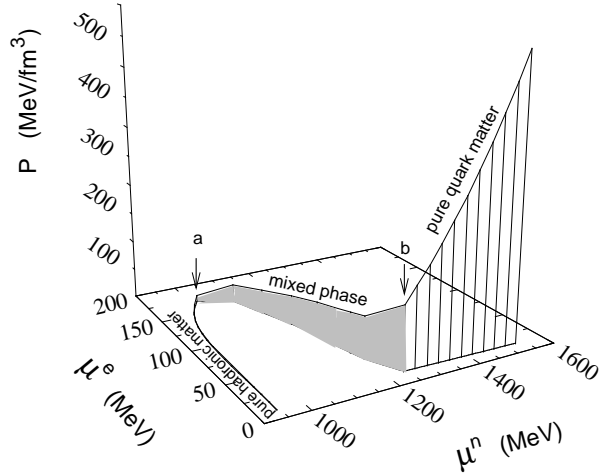,width=8.cm,angle=-90}
\caption[]{EoS of neutron star matter which accounts for quark
deconfinement \protect{\cite{weber99:book,weber99:topr}}.}
\label{fig:QuarkEoS}
\end{center}
\end{figure}
maximum mass of stable neutron stars in our theory is $2.2\,\msun$,
compact companions being more massive than that value are predicted to
be black holes. An example of such an object is the non-pulsating
X-ray binary Cyg~X--1, as its mass lies in the range $9 \lsim M /
\msun \lsim 15$.

\section{Evidence of quark matter in neutron stars}\label{sec:signal}

As pointed out in Section \ref{ssec:deconf}, the possibility of quark
deconfinement in the cores of neutron stars has already been suggested in the
1970's. However until recently no stringent observational signal has ever
\begin{figure}[tb] 
\begin{center}
\parbox[t]{6.5cm} {\leavevmode
\epsfig{figure=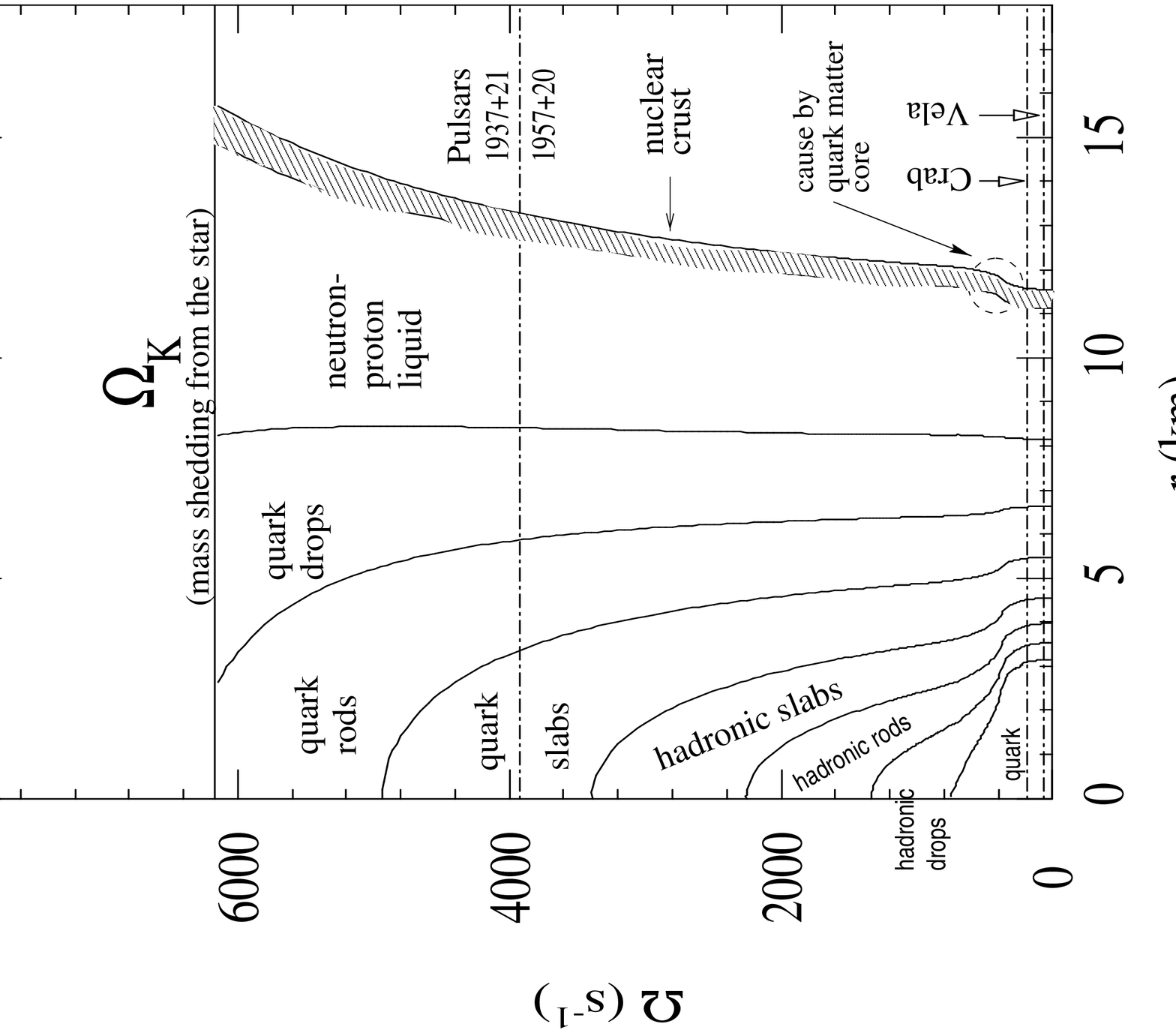,width=7.0cm,angle=-90}
{\caption[]{Frequency dependence of quark-hadron structure in
equatorial direction of a neutron star
\protect{\cite{weber99:book,weber99:topr}}.
\label{fig:freq.eq}}}}
\ \hskip 1.00cm   \
\parbox[t]{6.5cm} {\leavevmode
\epsfig{figure=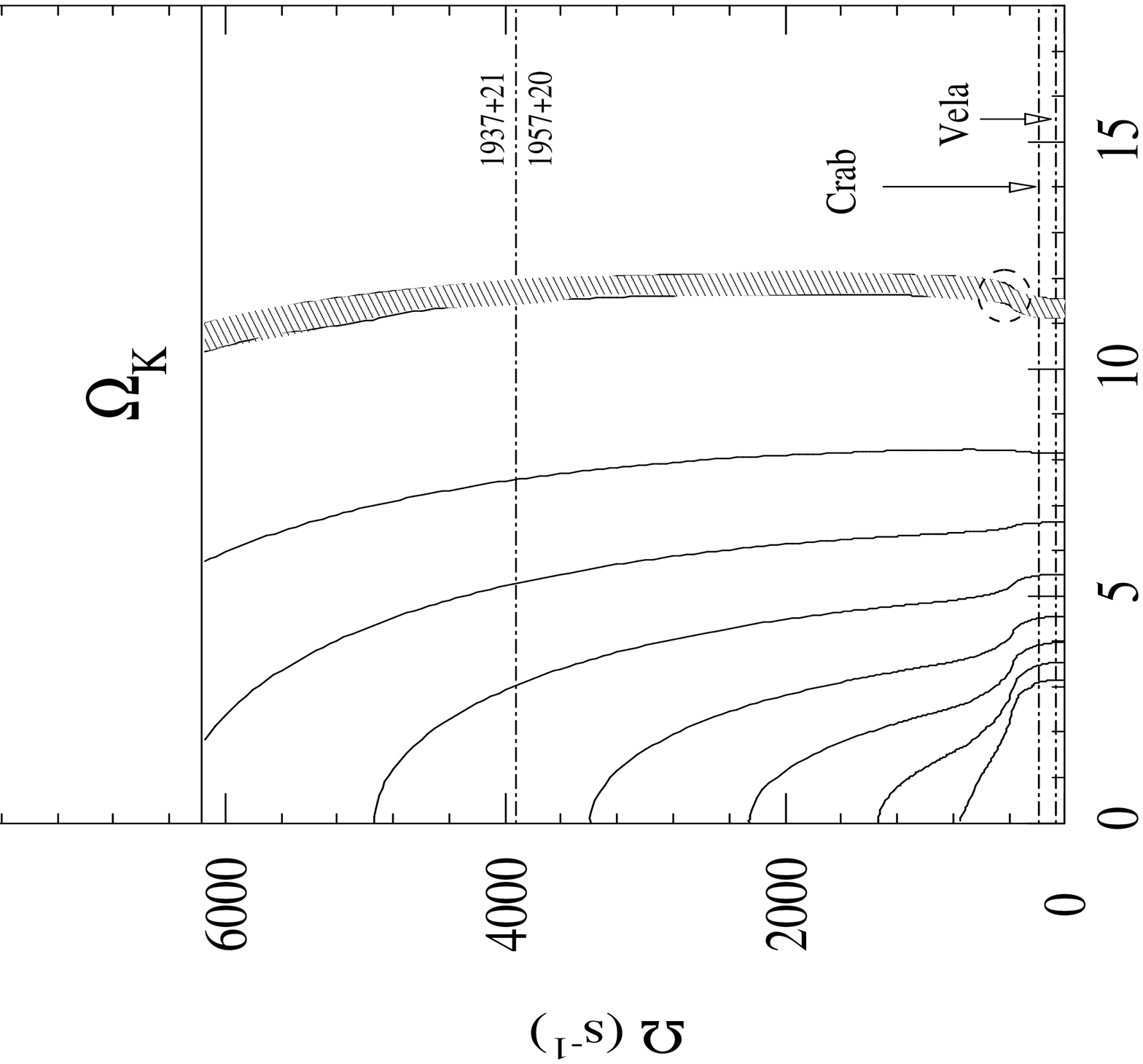,width=7.0cm,angle=-90} {\caption[]{Same as
Fig.~\protect{\ref{fig:freq.eq}}, but in star's polar direction
\protect{\cite{weber99:book,weber99:topr}}.}
\label{fig:freq.po}}}
\end{center}
\end{figure}
been proposed. This is so because whether or not the quark-hadron
phase transition occurs in neutron stars makes only little difference to
their static properties such as the range of possible masses, radii,
or even their limiting rotational periods. This, however, turns out to
be strikingly different for the timing structure of rotating neutron
stars (i.e. pulsars) that develop quark matter cores in the course of
spin-down, as I shall describe in this section.

A model of an EoS which accounts for quark deconfinement in neutron
star matter is shown in Fig.\ \ref{fig:QuarkEoS}.  One reads off that
the transition of confined hadronic matter to quark matter sets in at
about twice nuclear matter density (arrow labeled `a'), which leads to
a pronounced softening of the EoS.  Pure quark matter exists at
densities $\gsim 7 \epsilon_0$ (arrow labeled `b').

These structures manifest themselves inside rotating neutron stars as shown in
Figs.\ \ref{fig:freq.eq} and \ref{fig:freq.po} (non-rotating star mass is
$M=1.42\,\msun$).  The stars' baryon number is kept constant during spin-down
from the Kepler frequency to zero rotation, which describes the temporal
evolution of an isolated rotating neutron star.  Depending on frequency, the
star has an inner sphere of pure quark matter (labeled `quark' in
Fig.~\ref{fig:freq.eq}) surrounded by a few kilometers thick shell of mixed
phase of hadronic and quark matter arranged in a Coulomb lattice structure, and
this surrounded by a thin shell of hadronic liquid, itself with a thin crust of
heavy ions. The lattice structure of varying geometry may have dramatic effects
on pulsar observables including transport properties and the theory of glitches
\cite{glen97:book}. As the star spins up it becomes more and more deformed, and
the central density decreases. For some rotating neutron stars the
\begin{figure}[tb]
\begin{center}
\leavevmode
\epsfig{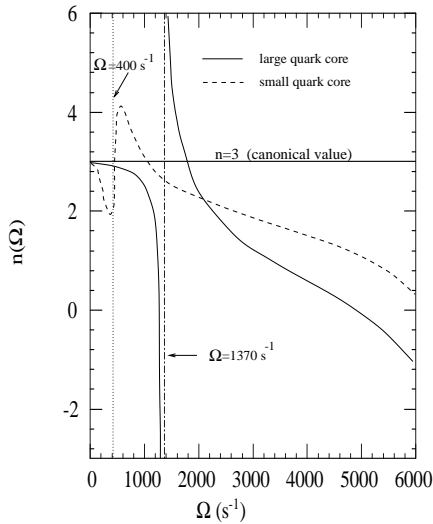}
\caption[]{Braking index as a function of rotational frequency 
\protect{\cite{weber99:topr}}.}
\label{fig:nvso}
\end{center}
\end{figure} mass and initial rotational frequency may be just such that the
central density rises from below to above the critical density for dissolution
of baryons into their quark constituents. This is accompanied by a sudden
shrinkage of the neutron star, which occurs for the present model star at
$\Omega \sim 400~\secm$. This effects the star's moment of inertia, $I$,
dramatically. Changes of $I$, in turn, reflect themselves in the braking index,
$n$, of a pulsar, which is given by $(I'\equiv {\rm d}I/{\rm d}\Omega,~
I''\equiv {\rm d}^2I/{\rm d}\Omega^2)$ \cite{weber99:book,glen97:a},
\begin{equation}  
  n(\Omega) \equiv \frac{\Omega\, \ddot{\Omega} }{\dot{\Omega}^2} = 3 - \frac{
    3 \, I^\prime \, \Omega + I^{\prime \prime} \, \Omega^2 } {2\, I + I^\prime
    \, \Omega} \, .
\label{eq:index}
\end{equation} One sees that the braking index depends explicitly and
implicitly on $\Omega$.  The right side reduces to the canonical constant $n=3$
only if $I$ is independent of frequency.  Figure \ref{fig:nvso} shows the
variation of $n$ with frequency for the star discussed just above (i.e.
Figs.~\ref{fig:freq.eq} and \ref{fig:freq.po}) as well as the sample star of
\cite{glen97:a}.  Because of the change in the moment of inertia driven by the
transition into quark matter, the braking index deviates dramatically from 3 at
the transition frequencies.  For the star of Figs.~\ref{fig:freq.eq} and
\ref{fig:freq.po}, this occurs at $\Omega \sim 400~\secm$ (dashed curve). The
solid curve shows the case of a strong deconfinement transition at $\Omega \sim
1370~\secm$ discussed in \cite{glen97:a}. Continuously connected intermediate
cases are possible too and were discussed in \cite{blaschke99:signal}.  Such
dramatic anomalies in $n(\Omega)$ are not known to be exhibited by conventional
neutron stars, because their moments of inertia increase smoothly with $\Omega$
\cite{weber99:book}. The radio astronomical observation of such an anomaly may
thus be interpreted as a signal for the development of quark-matter cores in
the centers of pulsars!

As a very important subject on this issue, we estimate the duration
\begin{figure}[tb]
\begin{center}
\leavevmode 
\epsfig{figure=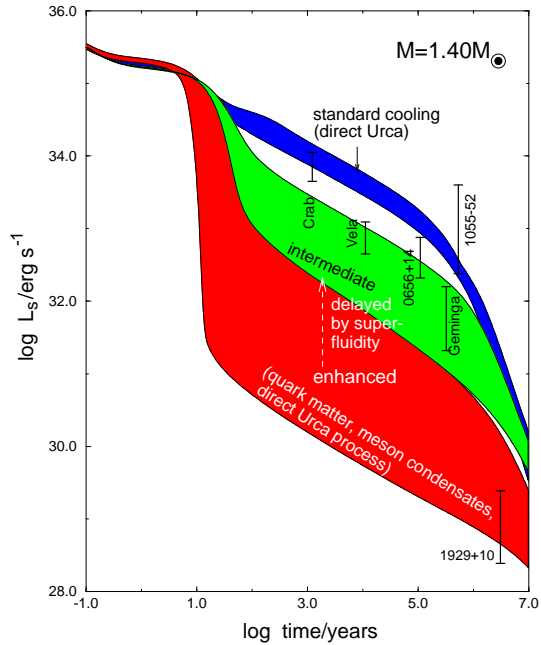,width=7.0cm}
\caption[]{Cooling behavior of a  $1.4\,\msun$ neutron star based on  
  competing assumptions about the behavior of superdense matter. Three distinct
  cooling scenarios, referred to as `standard', `intermediate', and `enhanced'
  (for details, see \protect{\cite{weber99:book}}), can be distinguished. The
  band-like structures reflects the uncertainties inherent in the underlying
  EoS (cf.\ Fig.\ \ref{fig:eos}).}
\label{fig:cool} 
\end{center}
\end{figure}
over which the braking index is anomalous. It can be estimated from
$\Delta T \simeq - \, {{\Delta\Omega}\over{\dot\Omega}} = {{\Delta
P}\over{\dot{P}}}$, where $\Delta \Omega$ is the frequency interval of
the anomaly.  For a millisecond pulsar whose period derivative is
typically $\dot{P}\simeq 10^{-19}$, one finds $\Delta T \simeq
10^8$~years.  The dipole age of such pulsars is about $10^9$~years.
So as a rough estimate we may expect that about 10\% of the $\sim 25$
solitary millisecond pulsars presently known, are in the transition
epoch and so could be signaling the ongoing process of quark
deconfinement, complementing the searches for the quark-gluon plasma
state at the terrestrial heavy-ion colliders.

\section{Cooling of neutron stars}

The predominant cooling mechanism of hot (temperatures of several
$\sim 10^{10}$~K) newly formed neutron stars immediately after
formation is neutrino emission, with an initial cooling time scale of
seconds. Already a few minutes after birth, the internal neutron star
temperature drops to $\sim 10^9$~K \cite{burrows86:a}. Photon emission
overtakes neutrinos only when the internal temperature has fallen to
$\sim 10^8$~K, with a corresponding surface temperature roughly two
orders of magnitude smaller. Neutrino cooling dominates for at least
the first $10^3$ years, and typically for much longer in standard
cooling (modified Urca) calculations.  Being sensitive to the adopted
nuclear \eosp, the neutron star mass, the assumed magnetic field
strength, the possible existence of superfluidity, meson condensates,
quark matter etc., theoretical cooling calculations, as summarized in
Fig.\ \ref{fig:cool}, too provide most valuable information about the
interior matter and neutron star structure. The stellar cooling tracks
in Fig.\ \ref{fig:cool} are computed for a broad collection of \eoss
\cite{weber99:book}, including those shown in Fig.\ \ref{fig:eos},
which account for the effects mentioned just above. Knowing the
thermal evolution of a neutron star also yields information about such
temperature-sensitive properties as transport coefficients, transition
to superfluid states, crust solidification, and internal pulsar
heating mechanisms such as frictional dissipation at the
crust-superfluid interfaces (for a recent overview,
see \cite{schaab99:b}).

\end{document}